# AN ALTERNATIVE CONSTRUCTION OF THE QUANTUM ACTION FOR SUPERGRAVITY

N. Djeghloul and M. Tahiri*

*Laboratoire de Physique Théorique,*

*Université d'Oran Es-Senia, 31100 Oran, Algeria*

**Abstract**

We develop a method to derive the onn-shell invariant quantum action of the supergravity in such a way that the quartic ghost interaction term is explicity determined. First, we reinvestigate the simple supergravity in terms of a principal superfibre bundle. This gives rise to the closed geometrical BRST algebra. Therefore we determine the open BRST algebra, which realizes the invariance of the classical action. Then, given a prescription to build the full quantum action, we obtain the quantum BRST algebra. Together with the constructed quantum action this allows us to recover the auxiliary fields and the invariant extension of the classical action.

* E-mail : ptmt@pt.tu-clausthal.de; also at Arnold Sommerfeld Institut für Mathematische Physik, Technische Universität Clausthal, 38678 Clausthal-Zellerfeld, Germany.

## 1. Introduction

Gauge theories are naturally described as geometrical theories over a principal fibre bundle. In this framework, the simple supergravity[1] has been discussed by considering the super-Poincaré group as structural group[2]. We recall that the only fields arising in this superfibre bundle approach are the vierbein, the gravitino and the spin connection. The latter is a nonpropagating field, since its equation of motion is algebraic. We also recall that the simple supergravity without auxiliary fields is a theory with an open algebra[1].

However, gauge theories whose gauge algebra is open (e.g. supersymmetric theories) and/or which are reducible (e.g. antisymmetric tensor gauge theories) can be quantized by the Batalin-Vilkovisky (BV) formalism, which provides on-shell Becchi-Rouet-Stora-Tyutin (BRST) invariant quantum actions[3,4]. This has been applied to the simple supergravity in Refs 5 and 6. In particular, in Ref. 6 this has been used to introduce the auxiliary fields.

On the other hand, it has been known that the superspace formulation of gauge theories provides a convenient framework for quantizing such theories[7]. In this formalism the BRST and anti-BRST transformations[3,8] can be derived by imposing horizontality conditions on the supercurvature associated to a superconnection on a superspace. The latter is obtaining by extending the space-time with two ordinary anticommuting coordinates. In this framework, the quantisation of topological antisymmetric tensor gauge theory, so-called BF theory[9], in four dimensions as well as of the simple supergravity has been performed by introducing auxiliary fields[10]. The superfield realization of general gauge theories has been discussed in Ref. 11.

Moreover, in the context of Yang-Mills theories, the horizontality conditions can be viewed as equations of motion by constructing a super-Lagrangian[12]. In Ref. 13 it has been shown how the horizontality conditions can be naturally incorporated in the geometrical structure of a principal superfibre bundle with a base space simply the space-time and a structure group the direct product of the gauge and the two-dimensional odd translation groups. There, the general formalism developed has been applied to the gravity in the vierbein formalism and the simple supergravity (see also Ref. 14 for the case of Yang-Mills theories and N=1, D=4 super-Yang-Mills theory). The same geometrical approach has been considered in order to realize the quantization of four dimensional BF theory which represents a model with a reducible symmetry[15]. Let us note that, in the case of the simple supergravity as discussed in Ref. 13, there are only the BRST and anti-BRST operators which have been determined. It is interesting to find out how we can construct the effective quantum action in relation to the superfibre bundle formalism.

Our purpose in this paper is basically twofold. On the one hand, we use a principal superfibre bundle which gives rise to the basic fields of the quantized simple supergravity, and allows us to determine in a rigorous way, contrary to what was done in Ref. 13, an off-shell nilpotent geometrical BRST operator. This, contrary to what was disscussed in Ref. 2, does not guarantee the invariance of the classical action of the simple supergravity. Then, to achieve the invariance, we are bound to build from the geometrical BRST operator an effective one. The obtained classical BRST transformations are nilpotent on-shell and agree with the standard results. On the other hand, we perform the construction of the on-shell BRST invariant quantum action. To this purpose, we work with the same spirit as in the BF theory[15]. We simply modify the classical BRST symmetry as well as the gauge-fixing action written as in Yang-Mills type theories. In addition, we show how the auxiliary fields in the simple supergravity arise, and how their BRST transformations can be determined.

The letter is structured as follow : in Sec. 2 we introduce the fields and obtain the geometrical BRST operator . We also discuss the invariance of the classical action and determine the nonclosure functions. In Sec. 3 we build the on-shell BRST invariant quantum action. We also discuss the transition to the auxiliary fields. Section 4 is devoted to concluding remarks.

Throughout this paper we take advantage of the Dirac matrices and use the Fierz rearrangement[1], in order to perform the calculations.

## 2.Classical BRST algebra

We start with the construction of a principal superfibre bundle $P(M,G_S)$ . The base space, $M$ , is the four-dimensional spacetime manifold and the structural group, $G_S$ , is the direct product of the (10,4)-dimensional super-Poincaré group (SP) and the (0,2)-dimensional translation Lie supergroup ($S^{0,2}$). We endow $P(M,G_S)$ with a superconnection $\Phi$, i.e. an even one-superform in the total space $P$ taking values in the Lie superalgebra of $G_S$ . The supercurvature $\Omega$ is, as usual, an even two-superform given by the structure equation, $\Omega = d\Phi + \frac{1}{2}[\Phi,\Phi]$ , where $d$ is the exterior superdifferential and [ , ] the graded Lie bracket. The supercurvature is a tensorial superform, i.e. we have $i(X)\,\Omega = 0$ , where $i(X)$ is the inner product along a vertical superfield $X$ in $P$ . It also satisfies the Bianchi identity, $d\Omega + [\Phi,\Omega] = 0$.

Knowing that $P$ is locally diffeomorphic to the product of $M$ and $G_S$ , we can define a local coordinate system $(Z^M)$ on $P$ as follows $(Z^M) = (x^m, y^i, q^a)$ , where $(x^m)_{m=1,\ldots,4}$ , $(y^i)_{i=1,\ldots,10+4}$ and $(q^a)_{a=1,2}$ are local coordinate systems on $M$, SP and $S^{0,2}$, respectively. Then, the superconnection $\Phi$ and its associated supercurvature $\Omega$ can be written as : $\Phi = dZ^M \Phi_M$ and $\Omega = \frac{1}{2} dZ^N \wedge dZ^M \Omega_{MN}$. Using these local expressions, the superfield components $\Omega_{MN}$ are determined by the structure equation, we have

$$\Omega_{MN} = \P_M \Phi_N - (-1)^{mn} \P_N \Phi_M + \left[\Phi_M, \Phi_N\right], \qquad (1)$$

where $m$ represents the Grassmann grade of $(Z^M)$. The Bianchi identity becomes

$$\tilde{D}_M \Omega_{NR} + (-1)^{m(r+n)} \tilde{D}_N \Omega_{RM} + (-1)^{r(m+n)} \tilde{D}_R \Omega_{MN} = 0 , \qquad (2)$$

where $\tilde{D}_M = \P_M + [\Phi_M, \ . \ ]$. We also obtain the following equations

$$\Omega_{ma} = \Omega_{ab} = 0 , \qquad (3)$$

$$\Omega_{mi} = \Omega_{ai} = \Omega_{ij} = 0 , \qquad (4)$$

according to the fact that $\P_i$ and $\P_a$ are vertical vectors superfields.

Moreover , let $(M_{ab}, P_a, Q_A)$ and $F_a$ be the generators of SP and $S^{0,2}$ , respectively. The components $\Omega_{MN}{}^a$ of the supercurvature associated to $F_a$ give the $S^{0,2}$-supertorsion, and according to Eq. (1), we have $\Omega_{MN}{}^a = \P_M \Phi_N{}^a - (-1)^{mn} \P_N \Phi_M{}^a$ . To remove the $S^{0,2}$-supertorsion dependence of the theory, we supplement Eqs. (3) and (4) with the constraint $\Omega_{mn}{}^a = 0$. Thus,

$$\Omega_{MN}{}^a = 0 . \qquad (5)$$

Therefore, $\Phi_M{}^a$ being pure gauge, we consider, hereafter, that the components $\Phi_M$ take values in the super-Poincaré algebra, i.e.

$$\Phi_M = \frac{1}{2} \Phi_M{}^{ab} M_{ab} + \Phi_M{}^a P_a + \Phi_M{}^A Q_A \ . \qquad (6)$$

Now, in order to give the interpretation of the fields occurring in the quantized simple supergravity and to determine their BRST transformations using the geometrical structure, described by Eqs. (1)-(6), of the superfibre bundle $P(M, G_S)$, we proceed as follows. First, let us note that it is the superconnection which represents, as usual, the gauge fields and their associated ghost and anti-ghost fields. This can be realized by imposing the following condition (see also Ref. 13 for a different approach)

$$\Phi_i = 0\,, \tag{7}$$

on the superconnection $\Phi$ in $P(M, G_S)$. Therefore, in view of Eq. (4), we must have also the conditions

$$\P_i \Phi_m = \P_i \Phi_a = 0\,, \tag{8}$$

which fulfill the consistency with Eq. (1). Thus the remaining superfield components $\Phi_m$ and $\Phi_a$ depend only on $x^m$ and $q^a$.

On the other hand, we assign to $q^1$ and $q^2$ the ghost numbers (-1) and (+1) respectively, and to $x^m, \Phi$ and the generators of SP the ghost number zero. According to Eq. (6) and $\Phi = dx^m \Phi_m + dq^a \Phi_a$ , these rules determine the ghost numbers of the superfields $(\Phi_m{}^{ab}, \Phi_m{}^a, \Phi_m{}^A, \Phi_1{}^{ab}, \Phi_2{}^{ab}, \Phi_1{}^a, \Phi_2{}^a, \Phi_1{}^A, \Phi_2{}^A)$ which are (0,0,0,+1,-1, +1,-1, +1,-1). This allows us to make the following identifications : $e_m{}^a = \Phi_m{}^a\big|$ is the vierbein, $w_m{}^{ab} = \Phi_m{}^{ab}\big|$ is the spin connection, $y_m = (y_m{}^A) = (\Phi_m{}^A\big|)$ is the gravitino field, $c^{ab} = \Phi_1{}^{ab}\big|$ is the Lorentz ghost, $c^{*ab} = \Phi_2{}^{ab}\big|$ is the anti-ghost of $c^{ab}$ , $c^a = \Phi_1{}^a\big|$ is the translation ghost, $c^{*a} = \Phi_2{}^a\big|$ is the anti-ghost of $c^a$, $c = (c^A) = (\Phi_1{}^A\big|)$ is the supersymmetric ghost, $c^* = (c^{*A}) = (\Phi_2{}^A\big|)$ is the anti-ghost of $c$ . We also realize the usual identifications : $\P_a S\big| = Q_a(S)\big|$, where $S$ is any superfield and $Q = Q_1$ $(Q^* = Q_2)$ is the geometrical BRST (anti-BRST) operator. The bar denotes the projection into the lowest component of the corresponding superfield.

Substituting Eqs. (3) and (6) into the structure equation (1), evaluating these at $q^a$ =0 and using the above identifications as well as the structure constants of SP[1], we obtain the following off-shell nilpotent geometrical BRST transformations:

$$Q e_m^a = D_m c^a - c^{ab} e_{mb} + \tfrac{1}{2} \bar{c} g^a y_m\,, \tag{9}$$

$$Q y_m = -\tfrac{1}{2} c^{ab} s_{ab} y_m + D_m c\,, \tag{10}$$

$$Q w_m^{ab} = D_m c^{ab}\,, \tag{11}$$

$$Q c^a = -c^{ad} c_d - \tfrac{1}{4} \bar{c} g^a c\,, \tag{12}$$

$$Qc = -\tfrac{1}{2}c^{ab}s_{ab}c, \tag{13}$$

$$Qc^{ab} = -c^{ad}c_d{}^{b}\ , \tag{14}$$

where $D_m c^a = \partial_m c^a + w_m{}^{ab}c_b$ , $D_m c = \partial_m c + \tfrac{1}{2}w_m{}^{ab}s_{ab}c$ and $D_m c^{ab} = \partial_m c^{ab} + w_m{}^{ad}c_d{}^{b} - w_m{}^{bd}c_d{}^{a}$. We also obtain the anti-BRST transformations, which can be derived from Eqs. (9)-(14) by the following mirror symmetry of the ghost numbers : $X \to X$ (for $X = e_m{}^a, w_m{}^{ab}, y_m$) and $X \to X^*$ (for $X = Q, c^a, c^{ab}, c$). However, the Stueckelberg, auxiliary fields $Qc^{*a} = h^a$, $Qc^{*ab} = h^{ab}, Qc^* = h$ $(Q^*c^a = h^{*a}, Q^*c^{ab} = h^{*ab}, Q^*c = h^*)$ in the BRST (anti-BRST) sector are related via the equation $\Omega_{12}\big| = 0$. We have

$$h^a + h^{*a} = -c^{ad}c_d^* - c^{*ad}c_d - \tfrac{1}{2}\bar{c}g^a c^*\ , \tag{15}$$

$$h^{ab} + h^{*ab} = -c^{ad}c^*{}_d{}^b - c^{*ad}c_d{}^b\ , \tag{16}$$

$$h + h^* = -\tfrac{1}{2}c^{ab}s_{ab}c^* - \tfrac{1}{2}c^{*ab}s_{ab}c\ . \tag{17}$$

Furthermore, due to Eqs. (3)-(5), the supercurvature can be written as: $\Omega = \tfrac{1}{2}dx^n \wedge dx^m \Omega_{mn}$, with

$$\Omega_{mn} = \tfrac{1}{2}\Omega_{mn}{}^{ab}M_{ab} + \Omega_{mn}{}^{a}P_a + \Omega_{mn}{}^{A}Q_A\ . \tag{18}$$

Together with Eqs. (1) and (6) this permit us to find

$$R_{mn}{}^{ab} = D_m w_n{}^{ab} - D_n w_m{}^{ab}\ , \tag{19}$$

$$T_{mn}{}^{a} = D_m e_n{}^{a} - D_n e_m{}^{a} - \tfrac{1}{2}\bar{y}_m g^a y_n\ , \tag{20}$$

$$S_{mn} = D_m y_n - D_n y_m\ , \tag{21}$$

where $R_{mn}{}^{ab} = \Omega_{mn}{}^{ab}\big|$, $T_{mn}{}^{a} = \Omega_{mn}{}^{a}\big|$ and $S_{mn} = (S_{mn}{}^{A}) = (\Omega_{mn}{}^{A}\big|)$ represent the Lorentz curvature, the supersymmetric torsion and the Fermi curvature, respectively ; $D_m w_n{}^{ab} = \partial_m w_n{}^{ab} + \tfrac{1}{2}(w_m{}^{ad}w_{nd}{}^{b} - w_m{}^{bd}w_{nd}{}^{a})$ , $D_m e_n{}^{a} = \partial_m e_n{}^{a} + w_m{}^{ab}e_{nb}$ and $D_m y_n = \partial_m y_n + \tfrac{1}{2}w_m{}^{ab}s_{ab}y_n$.

Let us now turn to discuss the invariance of the classical action $S_0$ of the simple supergravity under the above BRST transformations. Let us recall that $S_0$ is given by[1]

$$S_0 = \tfrac{1}{2}ee^m{}_a e^n{}_b R_{mn}{}^{ab} - \tfrac{1}{4}e^{mnrt}\bar{y}_m g_5 g_n S_{rt}\ , \tag{22}$$

where $e^m{}_a$ is the inverse of the vierbein defined by $e^m{}_a e_n{}^a = d_n^m$ and $e^m{}_a e_m{}^b = d_a^b$ , and $e = \det(e_m{}^a)$. In Eq. (22) and in what follows, the integration sign is omitted for simplicity.

To compute the $Q$-variation of $S_0$ , it is more convenient to determine the BRST transformations of $R_{mn}{}^{ab}$ and $S_{mn}$ through the Bianchi identity (2). We have

$$QR_{mn}{}^{ab} = -c^a{}_d R_{mn}{}^{db} + c^b{}_d R_{mn}{}^{da} \ , \tag{23}$$

$$QS_{mn} = -\tfrac{1}{2} c^{ab} s_{ab} S_{mn} + \tfrac{1}{2} R_{mn}{}^{ab} s_{ab} c \ \cdot \tag{24}$$

Together with Eqs. (9) and (10) these allow us to obtain, up to a total divergence,

$$QS_0 = \tfrac{1}{8} e^{mnrt} T_{mn}{}^a (e_{abcd} c^b R_{rt}{}^{cd} + \bar{c} g_5 g_a S_{rt}) - \tfrac{1}{8} e^{mnrt} c^a \bar{S}_{mn} g_5 g_a S_{rt} \ , \tag{25}$$

where $\bar{S}_{mn} = D_m \bar{y}_n - D_n \bar{y}_m$ and $D_m \bar{y}_n = \P_m \bar{y}_n - \tfrac{1}{2} w_m{}^{ab} \bar{y}_n s_{ab}$. In deriving Eq. (25), we have used (see Eq. (2))

$$\sum_{(mnr)} (\P_m R_{nr}{}^{ab} + w_m{}^{ad} R_{nrd}{}^b - w_m{}^{bd} R_{nrd}{}^a) = 0 \ , \tag{26}$$

$$\sum_{(mnr)} (\P_m S_{nr} + \tfrac{1}{2} w_m{}^{ab} s_{ab} S_{nr} - \tfrac{1}{2} R_{mn}{}^{ab} s_{ab} y_r) = 0 \ , \tag{27}$$

where $\sum_{(mnr)}$ means a cyclic sum over $m, n$ and $r$.

However, $T_{mn}{}^a = 0$ represents the equation of motion of $w_m{}^{ab}$. This can be solved for $w_m{}^{ab}$, one finds[1]

$$w_m{}^{ab} = \tfrac{1}{2} e^{nb} \Big[ \Big\{ \P_n e_m{}^a - \tfrac{1}{4} \bar{y}_n g^a y_m - (m \leftrightarrow n) \Big\} + \tfrac{1}{2} e_m{}^d e^{ra} \Big\{ \P_n e_{ra} - \tfrac{1}{4} \bar{y}_n g_d y_r - (n \leftrightarrow r) \Big\} \Big] - (a \leftrightarrow b) \ \cdot \tag{28}$$

Therefore using Eqs. (9) and (10), the BRST transformation of $w_m{}^{ab}$ as given in Eq. (11) must be replaced by

$$Qw_m{}^{ab} = D_m c^{ab} + d' w_m{}^{ab} \ ,$$

$$d' w_m{}^{ab} = \tfrac{1}{4} e^{na} \left( \bar{c} g^b S_{mn} + \tfrac{1}{2} e^{rb} \bar{c} g_m S_{rn} \right) + \tfrac{1}{2} c_d e^{na} e_{mf} \left( e^{rf} R_{rn}{}^{bd} + \tfrac{1}{2} e^{rb} R_{rn}{}^{fd} \right) - (a \leftrightarrow b) \ \cdot \tag{29}$$

The classical action of the simple supergravity is not invariant under the transformations (9), (10) and (29), since we have

$$QS_0 = -\tfrac{1}{8} e^{mnrt} c^a \bar{S}_{mn} g_5 g_a S_{rt} \ \cdot \tag{30}$$

We remark that the change $d'w_m^{ab}$ in the transformation of the spin connection induces a variation $d'S_0$, which is zero, since we can explicitly write, up to a total divergence,

$$d'S_0 = \tfrac{1}{4} e^{mnrt} e_{abcd} e_m^{\ a} T_{nr}^{\ \ b} d'w_t^{\ cd} \ . \tag{31}$$

Let us note that, by converting the ghosts into gauge parameters, the transformations (9), (10) and (29) become equivalent to those obtained in Ref. 2. There, the author fails to take into account properly the invariance of $S_0$ under such transformations, which leads to incorrect results and conclusions.

Let us also note that we shall modify the geometrical BRST transformations (9) and (10), in order to achieve the invariance of $S_0$. To this end, we remark that it is the Rarita - Schwinger action in $S_0$ which leads to a $Q$-variation of $S_0$ as in Eq. (30). So, we introduce a modified BRST operator $\tilde{Q}$ defined by $\tilde{Q} = Q + d$, with $de_m^{\ a} = 0$ and $dy_m \neq 0$. The $\tilde{Q}$-variation of $S_0$ now reads

$$\tilde{Q}S_0 = \tfrac{1}{2} e^{mnrt} \bar{S}_{mn} g_5 g_a \left( \tfrac{1}{4} c^a S_{rt} - e_r^{\ a} dy_t \right) , \tag{32}$$

and by a direct calculation, we find that

$$dy_m = e^n_{\ a} c^a S_{nm} , \tag{33}$$

satisfies $\tilde{Q}S_0 = 0$. Thus, the invariance of the classical action of the simple supergravity is guaranteed under the following transformations

$$\tilde{Q}e_m^{\ a} = c^n \P_n e_m^{\ a} + \P_m c^n e_n^{\ a} - c^{ab} e_{mb} + \tfrac{1}{2} \bar{c} g^a y_m , \tag{34}$$

$$\tilde{Q}y_m = c^n \P_n y_m + \P_m c^n y_n - \tfrac{1}{2} c^{ab} s_{ab} y_m + D_m c , \tag{35}$$

where we have introduced the following replacements

$$c^a \rightarrow c^m e_m^{\ a} , \tag{36}$$

$$c^{ab} \rightarrow c^{ab} + c^m w_m^{\ ab} , \tag{37}$$

$$c \rightarrow c + c^m y_m . \tag{38}$$

These fields redefinitions permit us, in particular, to put $\tilde{Q}w_m^{\ ab}$, by using Eqs. (28), (29) and (33), in a simple form. We have

$$\begin{aligned} \tilde{Q}w_m^{\ ab} &= c^n \P_n w_m^{\ ab} + \P_m c^n w_n^{\ ab} + D_m c^{ab} \\ &\quad + \tfrac{1}{4} e^{na} \bar{c} g^b S_{mn} - \tfrac{1}{4} e^{nb} \bar{c} g^a S_{mn} - \tfrac{1}{4} e^{na} e^{rb} \bar{c} g_m S_{nr} , \end{aligned} \tag{39}$$

where we have used the following Bianchi identity

$$\sum_{(mnr)} \left( e_{mb} R_{nr}{}^{ab} + \tfrac{1}{2} \bar{y}_m g^{a} S_{nr} \right) = 0 \ \cdot \tag{40}$$

We remark that the translation ghost $c^a$ has been replaced by the diffeomorphism (coordinate) ghost $c^m$ through the field redefinition (36); and the transformations (34), (35) and (39), by converting the ghost into gauge parameters, involve, as usual, the general coordinate, the local Lorentz and the supersymmetric transformations.

It is worth noting that the transformations (34) and (35) cannot be nilpotent; and a question arises is how to modify the transformations of the ghosts (12)-(14), in order to guarantee the nilpotency. We notice that $\tilde{Q}^2 e_m{}^a = (Q^2 + dQ) e_m{}^a$ ; and by using Eqs. (9)-(14), (29) and (33) as well as the field redefinitions (36)-(38), we obtain

$$\tilde{Q}^2 e_m{}^a = \P_n e_m{}^a dc^n + e_n{}^a \P_m dc^n - \left( dc^{ab} + \tfrac{1}{2} c^n c^r R_{nr}{}^{ab} + c^n d' w_n{}^{ab} \right) e_{mb}$$
$$- \tfrac{1}{2} \bar{y}_m g^{a} \left( dc + \tfrac{1}{2} c^n c^r S_{nr} \right) \cdot \tag{41}$$

So, we have $\tilde{Q}^2 e_m{}^a = 0$ , provided that

$$dc^n = 0 \ , \tag{42}$$

$$dc = -\tfrac{1}{2} c^n c^r S_{nr} \ , \tag{43}$$

$$dc^{ab} = -\tfrac{1}{2} c^n c^r R_{nr}{}^{ab} - c^n d' w_n{}^{ab} \ . \tag{44}$$

From this it follows that

$$\tilde{Q} c^m = c^n \P_n c^m - \tfrac{1}{4} \bar{c} g^m c \ , \tag{45}$$

$$\tilde{Q} c = c^n \P_n c - \tfrac{1}{2} c^{ab} s_{ab} c + \tfrac{1}{4} \bar{c} g^m c y_m \ , \tag{46}$$

$$\tilde{Q} c^{ab} = c^n \P_n c^{ab} - c^{ad} c_d{}^b + \tfrac{1}{4} \bar{c} g^m c w_m{}^{ab} \ . \tag{47}$$

Furthermore, according to Eqs. (34), (35), (39) and (45)-(47), we find

$$\tilde{Q}^2 y_m = V_{mn} \frac{dS_0}{d\bar{y}_n} \ , \tag{48}$$

$$\tilde{Q}^2 c^{ab} = Z_n{}^{ab} \frac{dS_0}{d\bar{y}_n} \ , \tag{49}$$

$$\tilde{Q}^2 X = 0 \ , \tag{50}$$

where $X$ represents the other fields, $\frac{dS_0}{d\bar{y}_m} = -\frac{1}{2e} e^{mnrt} g_5 g_n S_{rt}$ is the equation of motion of the gravitino, and

$$V_{mn} = \tfrac{1}{8}\bar{c} g^{\,r} c\left(\tfrac{1}{4} g_{mn} g_{\,r} - \tfrac{1}{2} e e_{mnrt} g_5 g^{\,t}\right)$$

$$+ \tfrac{1}{8}\bar{c} s^{\,rt} c\left(g_{mr} g_{nt} + \tfrac{1}{2} g_{mn} s_{\,rt} - \tfrac{1}{2} e e_{mnrt} g_5\right) , \tag{51}$$

$$Z_m^{\ ab} = \tfrac{1}{8}\bar{c} g_{\,m} s^{\,ab} g_5 c \quad \bar{c} g_5 , \tag{52}$$

where $g_{mn} = e_m^{\ a} e_{an}$. In particular, we note that, in deriving Eqs. (48) and (49), we have used the following identities

$$\sum_{(mnr)} g_{\,m} S_{nr} = e e_{mnrt} g_5 \frac{dS_0}{d\bar{y}_t} , \tag{53}$$

$$\tfrac{1}{4}\bar{c} g^{\,n} c S_{mn} + \tfrac{1}{4}\bar{c} g_{\,r} S_{mn} s^{\,nr} c - \tfrac{1}{8}\bar{c} g_{\,m} S_{nr} s^{\,nr} c = V_{mn} \frac{dS_0}{d\bar{y}_n} . \tag{54}$$

Thus, the classical BRST algebra as given in Eqs. (34), (35) and (45)-(47) is nilpotent on-shell. It has been obtained upon modifying the geometrical BRST algebra, in order to ensure the invariance of the classical action. Moreover, the nonclosure functions $V_{mn}$ and $Z_m^{\ ab}$ satisfy the following relations

$$V_{mn}^T = -C\, V_{mn} C^{-1} , \tag{55}$$

$$Z_m^{\ ab} = \tfrac{1}{2} e^{ra} e^{nb} \bar{c} g_{\,r} V_{nm} , \tag{56}$$

where $C$ is the charge conjugation matrix. The last equation may be also written as

$$V_{mn} g_{\,r} c = \tfrac{1}{8} e e_{mnrt} \bar{c} g^{\,t} c \; g_5 c . \tag{57}$$

## 3. Quantum BRST algebra

The classical BRST operator $\tilde{Q}$ of the simple supergravity is nilpotent on-shell. Hence, in order to build the full quantum action, we cannot simply apply the standard BRST quantization. This can be applied after introducing the auxiliary fields which are necessary to close the algebra. Without auxiliary fields, we can also perform the quantization, which leads to an on-shell BRST invariant quantum action involving quartic ghost interaction terms. Here, the quantum BRST transformations and the full quantum action coincide with those obtained upon elimination of the auxiliary fields after quantization (see Ref. 1 and references therein).

In the present section, we pursue to apply the method developed in the context of the BF theory where the symmetry is reducible[15], in order to realize the quantization of the simple supergravity where the classical algebra is open. To this purpose, we introduce a quantum BRST operator $\Delta$ defined by

$$\Delta = \tilde{Q} + \tilde{d} \ , \tag{58}$$

with $\tilde{d}X = 0$, where $X$ represents the fields satisfying $\tilde{Q}^2 X = 0$; and a quantum action $S_q$ written as

$$S_q = S_0 + S_{gf} \ , \tag{59}$$

where $S_{gf}$ is defined by

$$S_{gf} = (\tilde{Q} + x\tilde{d})\left(c^*_m G^m + c^*_{ab} G^{ab} + c^* G\right) \tag{60}$$

with $G^m$, $G^{ab}$ and $G$ being gauge conditions, which allow us to fix diffeomorphisms, local Lorentz transformations and supersymmetry transformations, respectively. In what follows, we choose the following standard gauge conditions for simplicity : $G^m = \partial_n (eg^{mn})$, $G^{ab} = \frac{e}{2}(e_m{}^a \tilde{e}^{mb} - e_m{}^b \tilde{e}^{ma})$ and $G = eg^m y_m$, where $g^{mn} = e^m{}_a e^{na}$ and $\tilde{e}_m{}^a$ is a background vierbein.

The operator $\tilde{d}$ and the numerical coefficient $x$ must be determined so that

$$\Delta S_q = 0 \ , \tag{61}$$

$$\Delta^2 = 0 \ , \tag{62}$$

where the last equation is written up to equations of motion at the quantum level.

Using Eqs. (48), (50) and (58), the $\Delta$-variation of $S_q$ reads

$$\Delta S_q = e\left(\tilde{d}\bar{y}_m + \bar{c}^* g^n V_{nm}\right)\frac{dS_0}{d\bar{y}_m} + xe\bar{c}^* g^m \tilde{d}^2 y_m$$

$$+ \tilde{d}\tilde{Q}\left(c^*_n \partial_m (eg^{mn}) + ec^*_{ab} e_m{}^a \tilde{e}^{mb}\right) + (\tilde{d}\tilde{Q} + x\tilde{Q}\tilde{d})\left(e\bar{c}^* g^m y_m\right) \ , \tag{63}$$

where we have used the fact that $\Delta S_0 = e\tilde{d}\bar{y}_m \frac{dS_0}{d\bar{y}_m}$. This is simply obtained after taking into account that the change $\tilde{d}w_m{}^{ab}$ in $\Delta w_m{}^{ab} = \tilde{Q}w_m{}^{ab} + \tilde{d}w_m{}^{ab}$ gives rise to a term in $\Delta S_0$, which is given by Eq. (31), where $d'$ is replaced by $\tilde{d}$.

The first term on the right hand side of Eq. (63) vanishes by taking

$$\tilde{d}\bar{y}_m = -\bar{c}^* g^n V_{nm} \ , \tag{64}$$

and in view of Eq. (55), we have

$$\tilde{d}y_m = -V_{mn} g^n c^* \ . \tag{65}$$

Therefore, the second term also vanishes, since $\tilde{d}^2 y_m = 0$, because of the definition of the quantum BRST operator $\Delta$. Moreover, the third term can be written as, (modulo a total divergence)

$$-ec^*{}_{ad}\ e_{nb}\tilde{e}^{nd}\tilde{d}c^{ab} \ ,$$

where we have used the fact that

$$\bar{c}g_n V_{mr} g^r j = 0 \ , \tag{66}$$

with $j$ being an arbitrary spinor. This simply follows from the following useful identity

$$V_{mn} g^n j = \tfrac{5}{64}\bar{c}g^n c\ g_{mn} g_n j \ , \tag{67}$$

by using Eq. (57). Finally, the last term on the right hand side of Eq.(63) can be put, modulo a total divergence, in the form

$$-\tfrac{e}{2}\bar{c}^* s_{ab} g^m y_m \tilde{d}c^{ab} - (x - \tfrac{1}{2})\tilde{Q}\left(e\bar{c}^* g^m V_{mn} g^n c^*\right) ,$$

which has been derived by using, in particular, the $\tilde{Q}$-variation of the nonclosure function $V_{mn}$ given by

$$\tilde{Q}V_{mn} = c^r \P_r V_{mn} + \P_m c^r V_{rn} + \P_n c^r V_{mr} - \tfrac{1}{2}c^{ab} s_{ab} V_{mn} + \tfrac{1}{2}c^{ab} V_{mn} s_{ab}$$
$$+\ \tfrac{1}{2}\bar{c}g^r y_r V_{mn} + \tfrac{1}{2}\bar{c}g^r y_m V_{rn} + \tfrac{1}{2}\bar{c}g^r y_n V_{mr} \ , \tag{68}$$

as well as Eqs. (55), (66) and (67).

So , the $\Delta$-variation of $S_q$ (63) acquires the form

$$\Delta S_q = -e\left(c^*_{ad} e_{nb}\tilde{e}^{nd} + \tfrac{1}{2}\bar{c}^* s_{ab} g^m y_m\right)\tilde{d}c^{ab} - (x - \tfrac{1}{2})\tilde{Q}\left(e\bar{c}^* g^m V_{mn} g^n c^*\right) . \tag{69}$$

Therefore, we can learn that the $\Delta$-invariance of $S_q$ is completely guaranteed by taking

$$\tilde{d}c^{ab} = 0 \ , \tag{70}$$

$$x = \tfrac{1}{2} \ . \tag{71}$$

However, after a similar straightforward calculation, we find

$$\Delta^2 y_m = V_{mn} \frac{dS_q}{d\bar{y}_n} + \tfrac{1}{2} \tilde{Z}_m{}^{ab} \frac{dS_q}{dc^{ab}} \ , \tag{72}$$

$$\Delta^2 c^{ab} = Z_m{}^{ab} \frac{dS_q}{d\bar{y}_m} \ , \tag{73}$$

$$\Delta^2 X = 0 \ , \tag{74}$$

where $X$ represents the other fields and $\tilde{Z}_m{}^{ab} = -C^{-1}\left(Z_m{}^{ab}\right)^T = \tfrac{1}{2} e^{ra} e^{nb} V_{mn} g_r c$.

We remark that the solutions (65) and (70) depend on the gauge conditions. In particular, we notice that $\tilde{d}c^{ab}$ can be derived from $\tilde{d}y_m$ by using Eq. (74) with $X = e_m{}^a, \ c$. In fact, we have

$$\tilde{d}c^{ab} = \tfrac{1}{2} e^{ma} e^{rb} \bar{c} g_m \tilde{d} y_n \ , \tag{75}$$

with $\bar{c} g_m \tilde{d} y_n = -\bar{c} g_n \tilde{d} y_m$. This vanishes for the solution (65), in view of Eq. (66).

Thus, we have constructed the full quantum action $S_q$ of the simple gravity

$$S_q = S_0 - \tfrac{e}{2} \bar{c}^* g^m V_{mn} g^n c^* + \tilde{Q}\left(c_n^* ¶_m (e g^{mn}) + e c_{ab}^* e_m{}^a \tilde{e}^{mb} + e \bar{c}^* g^m y_m\right) , \tag{76}$$

which is invariant under the on-shell nilpotent quantum BRST operator given by Eqs. (58), (65) and (70). The quantum action $S_q$ contains, as usual, four-ghost couplings described by the second term in the right hand side of Eq. (76).

Our final task is to show how we can introduce auxiliary fields so that the BRST algebra becomes closed. The aim of this construction is to give another possibility leading to the auxiliary fields in the simple supergravity, which avoids the application of the superspace formalism[1], the BV quantization procedure[6] and the BRST superspace approach[10]. To this end, looking at the quantum BRST transformation of the gravitino field, $\Delta y_m = \tilde{Q} y_m - V_{mn} g^n c^*$, we notice that the auxiliary fields shall be introduced via certain linearly independent combinations involving $c$ and $c^*$. Indeed, using Eq. (67), we also find the following identity

$$V_{\mu\nu}\gamma^{\nu}\chi^{*} = -\tfrac{1}{2}\Big(\tfrac{1}{4}\bar{\chi}\gamma_{\mu}\gamma_{5}\chi^{*}\gamma_{5}\chi - \tfrac{1}{3}\bar{\chi}\chi^{*}\gamma_{\mu}\chi + \tfrac{1}{3}\bar{\chi}\gamma_{5}\chi^{*}\gamma_{\mu}\gamma_{5}\chi - \tfrac{1}{12}\bar{\chi}\gamma_{\nu}\gamma_{5}\chi^{*}\gamma_{\mu}\gamma^{\nu}\gamma_{5}\chi\Big). \qquad (77)$$

Upon introducing

$$S = \bar{\chi}\chi^{*} \;,\quad P = \bar{\chi}\gamma_{5}\chi^{*} \;,\quad A_{\mu} = \tfrac{1}{4}\bar{\chi}\gamma_{\mu}\gamma_{5}\chi^{*} \;, \qquad (78)$$

the quantum BRST transformations can be written as

$$\begin{aligned} &\Delta e_{\mu}{}^{a} = \tilde{Q}e_{\mu}{}^{a} \;,\quad \Delta\psi_{\mu} = \tilde{Q}\psi_{\mu} + \tfrac{1}{2}\big(A_{\mu}\gamma_{5} + \gamma_{\mu}\eta\big)\chi \;,\\ &\Delta c^{\mu} = \tilde{Q}c^{\mu} \;,\quad \Delta\chi = \tilde{Q}\chi \;,\quad \Delta c^{ab} = \tilde{Q}c^{ab} + \tfrac{1}{2}\bar{\chi}\sigma^{ab}\eta\chi \;, \end{aligned} \qquad (79)$$

where $\eta = -\frac{1}{3}(S - P\gamma_{5} + A_{\nu}\gamma^{\nu}\gamma_{5})$. Furthermore, the quantum action (76) acquires the form

$$S_{q} = S_{0} - \tfrac{e}{3}\big(S^{2} - P^{2} + A_{\mu}A^{\mu}\big) + \Delta\Big(c^{*}_{\nu}\partial_{\mu}(eg^{\mu\nu}) + ec^{*}_{ab}e_{\mu}{}^{a}\tilde{e}^{\mu b} + e\bar{\chi}^{*}\gamma^{\mu}\psi_{\mu}\Big)\cdot \qquad (80)$$

Within this quantum action, Eq. (78) can be viewed as equations of motion by considering that $S$ , $P$ and $A_{\mu}$ become actually fields. They represent indeed the standard auxiliary fields in the simple supergravity. To this end, we shall determine their BRST transformations by imposing the off-shell nilpotency of the quantum BRST operator $\Delta$ as given in Eq. (79). Let us first remark that $\Delta^{2}X = 0$, where $X = c^{\mu}$ , $\chi$ , $e_{\mu}{}^{a}$ . For the last two fields, this is a direct consequence of the fact that the change $\tilde{\delta}c^{ab} = \frac{1}{2}\bar{\chi}\sigma^{ab}\eta\chi$ in $\Delta c^{ab}$ satisfies Eq. (75) with $\tilde{\delta}\psi_{\mu} = \frac{1}{2}(A_{\mu}\gamma_{5} + \gamma_{\mu}\eta)\chi$. On the other hand, acting $\Delta^{2}$ on $\Delta e_{\mu}{}^{a}$, we find that $\Delta^{2}c^{ab} = 0$ is automatically guaranteed provided that $\Delta^{2}\psi_{\mu} = 0$. This latter equation leads to

$$\Delta S = c^{\mu}\partial_{\mu}S - \tfrac{1}{4}\bar{\chi}\gamma_{\mu}R^{\mathrm{cov},\mu} \;, \qquad (81)$$

$$\Delta P = c^{\mu}\partial_{\mu}P - \tfrac{1}{4}\bar{\chi}\gamma_{5}\gamma_{\mu}R^{\mathrm{cov},\mu} \;, \qquad (82)$$

$$\Delta A_{b} = c^{\mu}\partial_{\mu}A_{b} - c_{bd}A^{d} - \tfrac{1}{4}e_{\mu b}\bar{\chi}\gamma_{5}(3R^{\mathrm{cov},\mu} - \gamma^{\mu}\gamma_{\nu}R^{\mathrm{cov},\nu}) \;, \qquad (83)$$

where $A_{b} = e^{\mu}{}_{b}A_{\mu}$ and $R^{\mathrm{cov},\mu} = \frac{1}{2e}\varepsilon^{\mu\nu\rho\tau}\gamma_{5}\gamma_{\nu}T_{\rho\tau}$ with $T_{\mu\nu} = \frac{1}{2}S_{\mu\nu} + \frac{1}{2}(A_{\mu}\gamma_{5} + \gamma_{\mu}\eta)\psi_{\nu} - (\mu \leftrightarrow \nu)$. In particular, we note that, in deriving Eqs. (81)-(83), the identity

$$\begin{aligned} &\bar{\chi}\gamma^{\nu}\chi\, T_{\mu\nu} + \bar{\chi}\gamma_{\rho}T_{\mu\nu}\,\sigma^{\nu\rho}\chi - \tfrac{1}{2}\bar{\chi}\gamma_{\mu}T_{\nu\rho}\,\sigma^{\nu\rho}\chi = \\ &\qquad \big(-\tfrac{1}{6}\bar{\chi}\gamma_{\nu}R^{\mathrm{cov},\nu}\big)\gamma_{\mu}\chi + \big(\tfrac{1}{6}\bar{\chi}\gamma_{5}\gamma_{\nu}R^{\mathrm{cov},\nu}\big)\gamma_{\mu}\gamma_{5}\chi \\ &\qquad + \Big\{-e_{\nu a}\bar{\chi}\gamma_{5}\big(3R^{\mathrm{cov},\nu} - \gamma^{\nu}\gamma_{\rho}R^{\mathrm{cov},\rho}\big)\Big\}\big(\tfrac{1}{6}\gamma_{\mu}\gamma^{a} - e_{\mu}{}^{a}\big)\gamma_{5}\chi \;, \end{aligned} \qquad (84)$$

has been used.

Furthermore, according to Eqs. (79) and (81)-(83), it is easy to check the $\Delta$-invariance of the extended action, $S_{inv} = S_0 - \frac{e}{3}(S^2 - P^2 + A_m A^m)$. So, starting from the on-shell BRST invariant quantum action, we have seen how it is possible to introduce the auxiliary fields, which close the BRST algebra and enable to recover the BRST invariant extension of the classical action[1].

## 4. Concluding remarks

We have performed the quantization of the simple supergravity starting from a principal superfibre bundle with superconnection in analogy to what is realized for the case of the BF theory[15].

The geometrical structure of the constructed superfibre bundle has been used to describe the fields occurring in the quantized simple supergravity and to determine the geometrical BRST operator. This, as we have seen, cannot represent a symmetry for the simple supergravity at the classical level, in contradiction to Ref. 2. In order to guarantee the invariance of the classical action, we have then constructed the on-shell classical BRST algebra. This has been realized, in particular, by modifying the geometrical BRST transformation of the gravitino field.

Starting from the classical BRST algebra, we have built the on-shell BRST invariant full quantum action of the simple supergravity, which involves, as usual, quartic ghost interaction terms. Furthermore, we have shown that the minimal set of auxiliary fields in the simple supergravity can be constructed upon introducing them as combinations of the supersymmetric ghost and its associated anti-ghost. The quantum BRST algebra and the quantized action have allowed us to recover their BRST transformations and the BRST invariant extension of the classical action[1].

Finally, we should mention that our aim in this letter has been to use a previous idea developped in Ref. 15, in order to realize the quantization of the simple supergravity as a model with open algebra. We have seen that it is possible and the standard results have been recovered. Let us note that in the case of BF theory[15] we can get a step further and introduce auxiliary fields working with the same spirit as discussed here. Furthermore, it would be interesting to find out how to construct on-shell BRST invariant quantum actions and to determine the structure of auxiliary fields in general gauge theories, independent of the underlying classical action, as generalization of the approach developped in this letter.

## Acknowledgments

M.T. would like to thank Prof. H.-D. Doebner for numerous discussion and constant support, and the Arnold Sommerfeld Institute for hospitality. He would also like to thank the Alexander von Humboldt Stiftung for the Georg Forester research fellowship.